\documentclass[prl,aps,twocolumn,groupedaddress,showpacs]{revtex4}
\usepackage{amssymb,amsmath,amsfonts,epsfig,graphicx,latexsym,bm,color}

\newcommand{\beq}{\begin{eqnarray}}
\newcommand{\eeq}{\end{eqnarray}}
\newcommand{\nn}{\nonumber}

\begin{document}
\title{Topologically induced avoided band crossing in an optical chequerboard lattice}
\author{Matthias \"{O}lschl\"{a}ger, Georg Wirth, Thorge Kock, and Andreas Hemmerich \footnote{e-mail: hemmerich@physnet.uni-hamburg.de} }
\affiliation{Institut f\"{u}r Laser-Physik, Universit\"{a}t Hamburg, Luruper Chaussee 149, 22761 Hamburg, Germany}
\date{\today}

\begin{abstract}
We report on the condensation of bosons in the 4th band of an optical chequerboard lattice providing a topologically induced avoided band crossing involving the 2nd, 3rd, and 4th Bloch bands. When the condensate is slowly tuned through the avoided crossing, accelerated band relaxation arises and the zero momentum approximately $C4$-invariant condensate wave function acquires finite momentum order and reduced $C2$ symmetry. For faster tuning Landau-Zener oscillations between different superfluid orders arise, which are used to characterize the avoided crossing.
\end{abstract}

\maketitle
Topologically non-trivial band structures provide the basis of intriguing forms of quantum matter such as high-temperature superconductors, unconventional superconductors in heavy-fermion compounds, quantum Hall systems or the recently discovered topological semi-metals in semiconductors with strong spin-orbit coupling \cite{Tok:00, Mae:04, Sto:99, Hsi:08}. Topological classification of band structures and band crossing points has been a topic of vibrant theoretical research recently \cite{Sun:08, Sun:09, Has:10, Qi:10, Kit:08, Kit:09}. The unique possibilities to prepare tailored periodic potentials in optical lattices \cite{Lew:07, Blo:08} have recently inspired numerous proposals to use them for clean simulations of topological matter \cite{Wu:08, XJLiu:10, Gol:10, Zha:11, VLiu:11, Sun:11}. In the lowest band artificial gauge fields may permit to engineer non-trivial topological features, however, at the cost of significant experimental complexity \cite{Dal:10}. In the higher bands of two- or three-dimensional non-separable band structures topological properties can be readily obtained by an appropriate choice of the lattice symmetries. Theoretical work has predicted that band relaxation can be kept moderate for bosonic atoms excited to higher bands \cite{Isa:05, Liu:06} and recently condensates in the $P$ and $F$-bands of a square optical lattice were reported \cite{Wir:11, Oel:11} and investigated theoretically \cite{Cai:11}. A complementary effort in exciton-polariton systems has led to a $D$-band condensate \cite{Kim:11}. 

In this work, we combine the tunability of optical lattice potentials with targeted condensate formation in an excited band to implement and investigate a topologically induced avoided band crossing. Extending techniques described previously \cite{Wir:11, Oel:11}, we produce an optical lattice potential with two classes of sites (denoted  as $\mathcal{A}$ and $\mathcal{B}$-sites in Fig.1(a)) arranged as the black and white fields of a chequerboard. In the $xy$-plane the optical potential with tunable parameters $V_0 \geq 0$ and $\Delta V$ is given by
\beq
\label{M.1}
V(x,y,\vec \epsilon) \,\equiv\, - \frac{V_0}{4} \,  | \, \eta \, \left(e^{i k x}  + \epsilon_{x} \,e^{-i k x} \right)  \qquad \qquad   \\ \nn 
+ \, e^{i \theta} \left(e^{i k y} + \epsilon_{y} \, e^{-i k y} \right) |^2 \,.
\eeq
with $\vec \epsilon \equiv (\eta,  \epsilon_{x},  \epsilon_{y})$ and carefully measured quantities $\eta = 0.98, \epsilon_{x} = 0.93, \epsilon_{y} = 0.87$ (accounting for imperfect optics in the experiment). The angle $\theta$ can be adjusted to better than $\pi/200$ by means of an interferometric optical set-up \cite{Hem:92}. The effective well depth difference is defined as $\Delta V(\theta) \equiv -V_0 \,\eta (1+\epsilon_{x}) (1+\epsilon_{y}) \cos(\theta)$, i.e., $\Delta V(\pi/2) = 0$. In the z-direction a weak harmonic potential is provided giving rise to elongated tubular lattice sites. 

\begin{figure}
\includegraphics[scale=0.55, angle=0, origin=c]{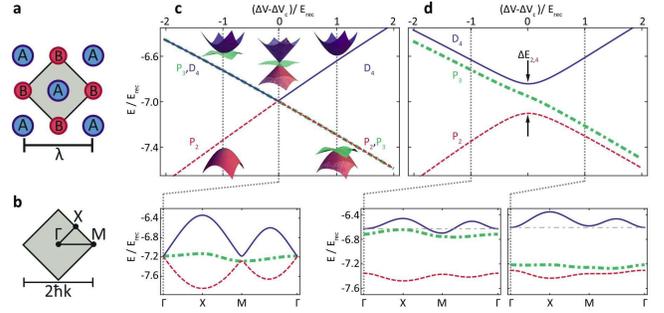}
\caption{\label{Fig.1} \textbf{a}, The lattice comprises two classes of sites with different well depth denoted by $\mathcal{A}$ and $\mathcal{B}$. The grey area denotes the Wigner Seitz unit cell. \textbf{b}, The first BZ with a path marked connecting the $\Gamma$-, $X$- and $M$-points, respectively. \textbf{c}, Energies of the 2nd ($B_2$), 3rd ($B_3$), and 4th ($B_4$) Bloch bands plotted versus $\Delta V$ for a lattice with perfect $C4$-symmetry for $V_0 = 7.0 E_{\textrm{rec}}$ and $\Delta V_{c} = 6.1 \,E_{\textrm{rec}}$. The detail on the lower edge shows a plot of these bands for $\Delta V = \Delta V_{c}$ within the 1st BZ along the trajectory illustrated in (b). \textbf{d}, The bands of \textbf{c} are replotted for a lattice with weakly broken $C4$-symmetry for $V_0 = 7.8 E_{\textrm{rec}}$ and $\Delta V_{c} = 6.1  \, E_{\textrm{rec}}$. The details on the lower edge show plots analogue to that shown in \textbf{c}, however, for $\Delta V = \Delta V_{c} \pm E_{\textrm{rec}}$. The (thin, grey) dashed dotted lines mark the energy of the 4th band at the $\Gamma$-point}
\end{figure}

We denote the Bloch bands as $B_n, n \in \{1,2,...\}$ ordered according to increasing energies. In Fig.1(c) the bands $B_2$, $B_3$ and $B_4$ at the $\Gamma$-point (cf. Fig.1(b)), derived from a two-dimensional (2D) band structure calculation for a lattice with $V_0 = 7.0 \, E_{\textrm{rec}}$ and perfect $C4$ symmetry (i.e., setting $\eta = \epsilon_{x} = \epsilon_{y} = 1$ in Eq.(1)), are plotted against $\Delta V$. Here, $E_{\textrm{rec}} = (\hbar k)^2/2m$ = recoil energy, $m$ = atomic mass, $k = 2\pi/\lambda$, and $\lambda = 1064\,$nm. The plot shows that at the critical value $\Delta V = \Delta V_{c}$ ($= 6.1\, E_{\textrm{rec}}$ for $V_0 = 7.0 \, E_{\textrm{rec}}$) all three bands become degenerate and the $B_2$- and the $B_4$-band form a Dirac cone in the centre of the first Brillouin zone (BZ) with the locally flat $B_3$-band intersecting its origin. This can be seen in the detail on the lower edge of Fig.1(c) showing the involved bands within the first BZ along a trajectory connecting the points $\Gamma$, $X$, $M$, $\Gamma$ of Fig.1(b). For $\Delta V >\Delta V_{c}$ ($\Delta V < \Delta V_{c}$) the $B_4$-band ($B_2$-band) separates, whereas $B_2$ and $B_3$ ($B_3$ and $B_4$) remain degenerate, thus forming a topologically protected quadratic band crossing point (TQB) \cite{Sun:09} at the centre of the first BZ. This structure is robust against changes of $\Delta V$ and $V_{0}$ as long as the $C4$ symmetry is sustained.

In our experimental realization, due to technical imperfections accounted for by the value of $\vec \epsilon$ specified below Eq.(1), the $C4$ symmetry of the lattice is weakly broken. The lattice potential may be decomposed into a main contribution with perfectly $C4$-invariant unit cell and a small perturbation with no rotation symmetry, which acts to lift the degeneracies in Fig.1(c), thus leading to an avoided band crossing with an energy gap $\Delta E_{2,4}$ on the order of a small fraction of $E_{\textrm{rec}}$, as shown in Fig.1(d). A band calculation shows that $\partial \Delta E_{2,4}/\partial V_0 < 0.02$, i.e., the gap $\Delta E_{2,4} \approx 0.26 \, E_{\textrm{rec}}$ is nearly independent of $V_0$. On the left side of the anti-crossing ($\Delta V < \Delta V_{c}$) the bands $B_3$ and $B_4$ rapidly approach an approximately constant separation $\Delta E_{3,4} \approx 0.13 \,E_{\textrm{rec}}$, which is nearly independent of $V_0$ ($\partial \Delta E_{3,4}/\partial V_0 < 0.002$). Within the 1st BZ the energy $\Delta E_{3,4}$ appears as the gap introduced into the TQB at the $\Gamma$-point. An analogue $V_0$-independent gap $\Delta E_{2,3}$ arises for the bands $B_2$ and $B_3$ on the right side of the anti-crossing ($\Delta V > \Delta V_{c}$). The energy gaps have been calculated according to the definition $\Delta E_{n,m}(V_0) \equiv \textrm{Min}_{\Delta V \in [\Delta V_{c}-2E_{\textrm{rec}},\Delta V_{c}+2E_{\textrm{rec}}]} [E_m(\Delta V,V_0)-E_n(\Delta V,V_0)]$ with $E_n(\Delta V,V_0)$ denoting the energy of the nth band. The robustness of these gaps (with sizes on the order of ten nanokelvin) against changes of $V_0$ and $\Delta V$ indicate that, despite the broken $C_4$-symmetry, some topological character of the bands appears to be preserved. A theoretical method (extending the Berry flux concept \cite{Hal:04, Sun:09}) to quantify this residual robustness is yet to be conceived. The details on the lower edge of Fig.1(d) show $\Gamma X M$-trajectory plots of the bands within the first BZ away from the avoided crossing at $\Delta V = \Delta V_{c} \pm E_{\textrm{rec}}$.

\begin{figure}
\includegraphics[scale=0.6, angle=0, origin=c]{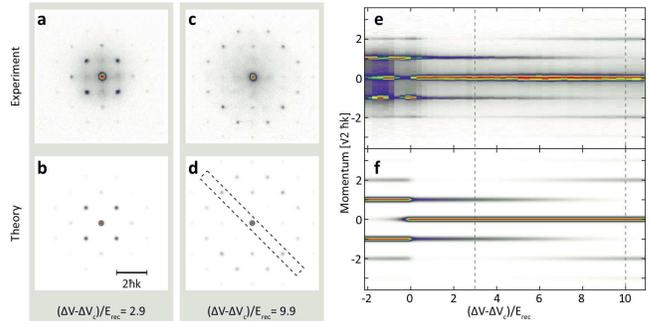}
\caption{\label{Fig.2} Momentum spectra of the $B_4$-band condensate are shown in \textbf{a} and \textbf{b} for $(\Delta V - \Delta V_{c})/ E_{\textrm{rec}} = 2.9$ and in \textbf{c} and \textbf{d} for $(\Delta V - \Delta V_{c})/ E_{\textrm{rec}} = 9.9$, respectively. The experimental results in \textbf{a} and \textbf{c} are compared to calculations in \textbf{b} and \textbf{d}. The $\Delta V$ dependence of the populations of the Bragg peaks identified in the dashed black rectangle in (d) is shown in (e) (experimental observations) and (f) (calculations).}
\end{figure}

As the initial step in our experimental protocol we produce a sample of about $10^5$ rubidium atoms ($^{87}$Rb) in the $F=2, m_F=2$ electronic ground state condensed in the global minimum of the $B_4$-band for a value $\Delta V = \Delta V_{i} \equiv 9.0 \,E_{\textrm{rec}} = \Delta V_{c} + 2.9 \,E_{\textrm{rec}}$ located on the right hand side of the avoided crossing in Fig.1(d), i.e., where the $B_4$-band is well separated from all other bands. This value is chosen in order to maximize the decoherence time of the condensate fraction to about $26$ ms, which corresponds to a lifetime of the $B_4$-band poplation of $45$ ms. For condensate formation (following Refs. \cite{Wir:11, Oel:11}) a deep ground state lattice ($V_{0} = 15 E_{\textrm{rec}}$) is prepared with $\Delta V \ll 0$ such that only the deeper $\mathcal{B}$-sites are occupied and tunneling is suppressed. Then, $\Delta V$ is switched to $\Delta V_{i}$, rendering $\mathcal{B}$-sites more shallow than $\mathcal{A}$-sites, and $V_{0}$ is reduced to $7.8\,E_{\textrm{rec}}$ in order to allow for tunneling. Within $10$ ms the system condenses into the minimum of the $B_4$-band and we may subsequently tune $\Delta V$ to any final value of interest. 

Fig.2 shows that in a wide range of values $\Delta V$ we observe momentum spectra with sharp Bragg maxima, resulting from the coherent condensate fraction residing at the $\Gamma$-point of the $B_4$-band. The panels in (a) and (c) show the experimental observations for $\Delta V = \Delta V_{i} = \Delta V_{c} + 2.9 \,E_{\textrm{rec}}$ and $\Delta V =  \Delta V_{c} + 9.9 \, E_{\textrm{rec}}$, respectively. The panels below in (b) and (d) present corresponding calculations based upon single particle Bloch-functions using the potential of Eq.(1), showing good agreement with respect to the relative sizes of the higher order Bragg peaks. For $\Delta V > \Delta V_{c}$, the Bloch-function corresponding to the $\Gamma$-point of the $B_4$-band comprises local $1s$-orbits in the shallow wells and local $3s$-orbits (with one radial node) in the deep wells. For values $\Delta V$ in the vicinity of $\Delta V_{i}$ (as in (a) and (b)) the $1s$-orbits hold most of the population, while the $3s$-orbits are only marginally populated. Hence, the envelope of the momentum spectrum is mainly due to the Fourier transform of the comparatively delocalized Wannier-function of the $1s$-orbit, such that only a few higher order Bragg peaks appear. Larger values of $\Delta V$ yield increased population of the $3s$-orbits. Because of their larger momentum components, higher order Bragg peaks become increasingly visible. This is more systematically studied in panel (e), which shows the $\Delta V$-dependence of the populations of the Bragg peaks on the descending diagonal in (d) indicated by the grey dashed box. The corresponding calculation in (f) shows striking agreement with the observations. For values of $\Delta V > \Delta V_{c}$ the $B_4$-band is $C4$-invariant to very good approximation with its global minimum at the $\Gamma$-point. In this case the zero momentum Bragg peak is by far the most populated. If we tune $\Delta V$ to values below $\Delta V_{c}$, thus passing the avoided band crossing, significant oscillations of the Bragg peak populations arise, which are discussed below. 

\begin{figure}
\includegraphics[scale=0.55, angle=0, origin=c]{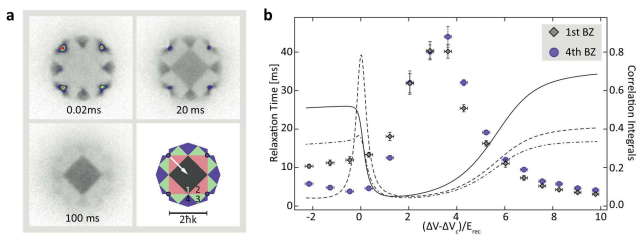}
\caption{\label{Fig.3} Band mapping plots recorded for $(\Delta V - \Delta V_{c})/ E_{\textrm{rec}} = 4.5$ and holding times $20\, \mu$s, $20\,$ms, $100\,$ms. In the lower right corner a map of the theoretical BZs 1,2,3,4 is plotted. The black circles indicate the observed condensation points at the energy minima of the $B_4$-band, which connect to the centre of the 1st zone via reciprocal lattice vectors (white arrow). \textbf{b}, The decay time of the 4th (blue disks) and the refilling time of the 1st band (black diamonds) are plotted versus $\Delta V$. The black line plots show the scaled overlap integrals $I_{1}$ (solid), $I_{2}$ (dashed), $I_{3}$ (dashed dotted) defined in Eq.(2).}
\end{figure}

In Fig.3 we study the collision induced relaxation of the $B_4$-band population. After condensate formation, $\Delta V$ is adjusted to some desired value within $200 \mu$s and the atoms are held in the lattice for a variable time. Finally, the lattice potential is adiabatically turned off, and the atoms are allowed to expand for 30 ms. This provides us with images of momentum space, in which the population of the nth band is mapped into the nth BZ. In (a) an example is shown for $(\Delta V - \Delta V_{c})/ E_{\textrm{rec}} = 4.5$, where the effect of the avoided crossing is not relevant. Initially (for 20 $\mu$s holding time), mainly the 4th BZ is populated, with a significant fraction of the atoms residing at the condensation points $(\pm \hbar k,\pm \hbar k)$. As seen in the pictures for larger holding times, the population of the 4th band directly decays into the 1st band (with approximately exponential time dependence), while the 2nd and 3rd BZs are not markedly involved. Only in the vicinity of the avoided crossing, a more complex decay dynamics arises with the 2nd and 3rd bands initially accumulating significant populations before finally the 1st BZ is refilled. This becomes visible in (b) (showing the relaxation times for the 4th and the 1st bands versus $\Delta V$), where close to the anti-crossing ($\Delta V \approx \Delta V_{c}$) the decay of the 4th band (blue disks) is faster than the refilling of the 1st band (black diamonds). The plot shows a pronounced resonance around $\Delta V = \Delta V_{i}$ with notably long lifetimes above 40 ms. This may be qualitatively explained by the observation that around $\Delta V_{i}$ most of the atomic population resides in the local $1s$-orbits of the shallow wells, where it is protected from collisional decay, because locally no state with lower energy is available. This assertion is supported by plotting the integrals 
\beq \label{CollisionIntegrals}
I_{n}(\Delta V) = \frac{\int_{\diamond}d^2r\, \rho_{4} \rho_{n}} {  \sqrt{ \int_{\diamond} d^2r \, \rho_{4}^2 \, \int_{\diamond} d^2r\,\rho_{n}^2   }    }
\eeq
where $\diamond$ denotes the unit cell in configuration space, $\rho_{n}\equiv  |\phi_{n}|^2$ and $\phi_{n}, n\in \{1,2,3,4\}$ denotes the Bloch-function of the nth band for zero quasi-momentum calculated for the optical potential in Eq.(1). These integrals measure the spatial correlations of the particle densities in the 4th and the nth bands at the $\Gamma$-point. As shown in (b) the maximal lifetime of the 4th band arises, where these overlap integrals are small. The observation in (a) that, apart from the vicinity of the avoided crossing, direct decay to the 1st band appears as the dominant process, corresponds to the fact that $I_{1}$ displays the largest decrease near the relaxation time maximum. 

\begin{figure}
\includegraphics[scale=0.63, angle=0, origin=c]{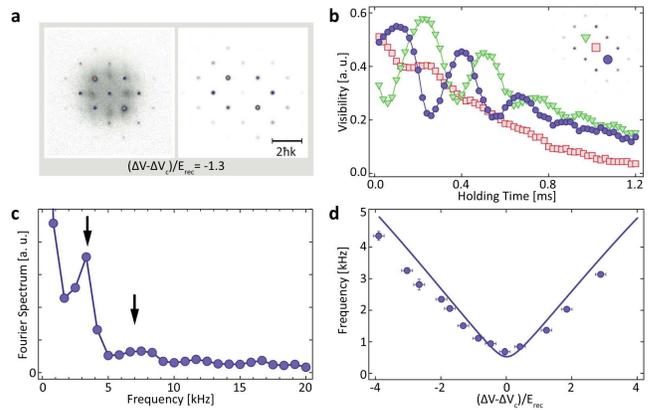}
\caption{\label{Fig.4} \textbf{a}, Momentum spectra (left = observation, right = calculation) on the left side of the anti-crossing ($\Delta V = \Delta V_{c} - 1.3\,E_{\textrm{rec}}$). A sufficiently large gap ($\approx 0.8\, E_{\textrm{rec}}$) was adjusted, in order to prevent Landau-Zener dynamics. \textbf{b}, Landau-Zener dynamics for a gap of $\approx 0.26 E_{\textrm{rec}}$. The visibility of the $\pm (1,-1) \hbar k$-Bragg peaks (blue disks and green triangles) and that of the $(0,0) \hbar k$-peak (red squares) are plotted versus the holding time after rapid tuning over the anti-crossing to $(\Delta V -  \Delta V_{c}) / E_{\textrm{rec}} = -3.0$. \textbf{c}, Fourier spectrum of the oscillating $(-1,1) \hbar k$-peak with first and second harmonic components indicated by black arrows. \textbf{d}, Blue disks denote the first harmonic frequencies derived from plots as in (b). The solid line shows the calculated energy difference between the 4th and the 2nd band.}
\end{figure}

As already indicated in Fig.2, when we drive the condensate across the avoided crossing at $\Delta V = \Delta V_{c}$, we observe momentum spectra with oscillating populations of the Bragg maxima. For small band gaps of less than about half a recoil energy, the rapid band relaxation occurring near the avoided crossing (cf. Fig.3(b)) limits us to ramping times below a millisecond, such that we operate in the non-adiabatic regime, where significant Landau-Zener oscillations are to be expected. To access the adiabatic regime, we had to broaden the band gap to $0.8\,E_{\textrm{rec}}$ by introducing larger deviations from $C4$-symmetry. In this case, we can maintain most atoms in the 4th band over the entire anti-crossing. As the anti-crossing is passed (with decreasing $\Delta V$, i.e, from right to left in Fig.1(d)) the condensate wave function undergoes a dramatic change. While on the right side of the anti-crossing a $C4$-invariant momentum spectrum with a leading zero momentum peak is observed, similar to that shown in Fig.2(a), on the left side of the anti-crossing the spectrum acquires finite momentum character and reduced $C2$ symmetry with the leading Bragg-orders arising at $\pm (1,-1) \hbar k$, as shown in Fig. 4(a) for $\Delta V -  \Delta V_{c}) / E_{\textrm{rec}} = -1.3$. The rotational symmetries seen in the momentum spectra reflect those of the $B_4$-band calculated for the respective values of $\Delta V$. Correspondingly, on the right side of the anti-crossing the calculated zero momentum Bloch-function displays approximate $C4$-symmetry with local $3s$-orbits and local $1s$-orbits in the deep and shallow wells, respectively, while on the left side, the deep wells comprise local $2p_{x-y}$-orbits aligned along the $(1,-1)$-direction.

The non-adiabatic case was studied for $\Delta E_{2,4} \approx 0.26 \, E_{\textrm{rec}}$ corresponding to 25 nK. After preparation of the condensate at $\Delta V = \Delta V_{i}$, within $400\,\mu$s we tune to values $\Delta V < \Delta V_{c}$ on the left side of the avoided crossing, and record the time evolution of the momentum spectrum. An example for $(\Delta V -  \Delta V_{c}) / E_{\textrm{rec}} = -3.0$ is shown in Fig.4(b), where the visibility \cite{visibility} of the $\pm (1,-1) \hbar k$-Bragg peaks (blue disks and green triangles) and that of the $(0,0) \hbar k$-peak (red squares) are plotted versus the holding time after the jump over the anti-crossing. We observe significant oscillations, which can be qualitatively modeled as a beat between the single-particle Bloch-functions of the 4th and 2nd band at the $\Gamma$-point. A closer inspection shows, that the frequency spectra of these oscillations comprise small second harmonic components as is illustrated for the $(-1,1) \hbar k$-peak in Fig.4(c), which is expected as a result of the non-linearity introduced by collisional interactions. The second harmonic contributions give rise to slightly increased (decreased) curvatures in the minima (maxima) of the oscillations in Fig.4(b). Collisional relaxation is also responsible for the observed decay of the visibility. If an additional lattice potential along the z-direction is applied in order to increase the collision energy per particle, we find correspondingly decreased decay times. Evaluation of curves similar as in Fig.4(b) for variable $\Delta V$ yields the blue disks in Fig.4(d). The solid trace repeats the energy difference between the 4th and the 2nd band from Fig.1(d) using the measured value of $\vec \epsilon$ specified below Eq.(1). Despite neglecting collisions, the calculations without use of fitted parameters well approximate the observations. The small deviations of the observed frequencies towards values slightly below the single particle calculations cannot be reduced by choosing different values for $\eta, \epsilon_{x}, \epsilon_{y}$ in Eq.(1), but rather indicate the effect of collisions. A simplified non-linear Landau-Zener model with the three zero momentum Bloch functions of the band crossing in Fig.1(c) as basis modes, a $3 \times 3$ coupling matrix adjusted to reproduce the single-particle anti-crossing in Fig.1(d), and a collision matrix accounting for the collision processes among the three basis modes confirms the observed trend, however, without yielding quantitative agreement. 

\begin{acknowledgments}
This work was partially supported by DFG (He2334/13-1, SFB 925, GrK 1355) and the Excellence cluster "Frontiers in Quantum Photon Science". We are grateful to C. Morais Smith and L. Mathey for helpful discussions.
\end{acknowledgments}


\begin{thebibliography}{23}
\bibitem{visibility}
The visibility is defined as $(n_{\textrm{sig}} - n_{\textrm{ref}})/(n_{\textrm{sig}} + n_{\textrm{ref}})$, where $n_{\textrm{sig}}$ denote the number of atoms in a disk-shaped area centered at the Bragg-peak of interest and $n_{\textrm{ref}}$ is the number of atoms in a ring-shaped domain surrounding this disk with the same area.
\end{thebibliography}
\end{document}